\documentclass[fleqn,usenatbib]{mnras}

\usepackage{newtxtext,newtxmath}
\usepackage[T1]{fontenc}
\usepackage{eso-pic}% http://ctan.org/pkg/eso-pic

\DeclareRobustCommand{\VAN}[3]{#2}
\let\VANthebibliography\thebibliography
\def\thebibliography{\DeclareRobustCommand{\VAN}[3]{##3}\VANthebibliography}

\usepackage{graphicx}	% Including figure files
\usepackage{amsmath}	% Advanced maths commands
\usepackage{ctable}

\newcommand{\DESVYR}{DES-SN5YR}
\newcommand{\DESVYRstar}{DES-SN5YR$^{\ast}$}

\newcommand{\DESIIIYR}{DES-SN3YR}
\newcommand{\pplus}{Pantheon+}
\newcommand{\SNANA}{{\tt SNANA}}

\newcommand{\URLBBC}{\url{https://github.com/RickKessler/SNANA/blob/master/src/SALT2mu.c}}
\newcommand{\Om}{\Omega_{\rm M}}

\newcommand{\mubias}{\mu_{\rm bias}}

\newcommand{\dmuOffset}{\Delta\mu_{\rm offset}}

\newcommand{\mxstd}{m_x^{\mathrm{std}}}

\title[Comparing DES-SN5YR and Pantheon+ analyses]{Comparing the \DESVYR\ and \pplus\ SN cosmology analyses: \\Investigation based on ``Evolving Dark Energy or Supernovae systematics?''}

\author[Vincenzi et al.]{\parbox{\textwidth}{M.~Vincenzi,$^{1}$\thanks{E-mail: maria.vincenzi@physics.ox.ac.uk}
R.~Kessler,$^{2,3}$
P.~Shah,$^{4}$
J.~Lee,$^{5}$
T.~M.~Davis,$^{6}$
D.~Scolnic,$^{7}$
P.~Armstrong,$^{12}$
D.~Brout,$^{8}$
R.~Camilleri,$^{6}$
R.~Chen,$^{7}$
L.~Galbany,$^{9,10}$
C.~Lidman,$^{11,12}$
A.~M\"oller,$^{13}$
B.~Popovic,$^{14}$
B.~Rose,$^{15}$
M.~Sako,$^{5}$
B.~O.~S\'anchez,$^{16}$
M.~Smith,$^{17}$
M.~Sullivan,$^{18}$
P.~Wiseman,$^{18}$
T.~M.~C.~Abbott,$^{19}$
M.~Aguena,$^{20}$
S.~Allam,$^{21}$
F.~Andrade-Oliveira,$^{22}$
S.~Bocquet,$^{23}$
D.~Brooks,$^{4}$
A.~Carnero~Rosell,$^{24,20,25}$
J.~Carretero,$^{26}$
L.~N.~da Costa,$^{20}$
M.~E.~S.~Pereira,$^{27}$
H.~T.~Diehl,$^{21}$
P.~Doel,$^{4}$
S.~Everett,$^{28}$
B.~Flaugher,$^{21}$
J.~Frieman,$^{21,3}$
J.~Garc\'ia-Bellido,$^{29}$
E.~Gaztanaga,$^{10,30,9}$
D.~Gruen,$^{23}$
R.~A.~Gruendl,$^{31,32}$
G.~Gutierrez,$^{21}$
S.~R.~Hinton,$^{6}$
D.~L.~Hollowood,$^{33}$
K.~Honscheid,$^{34,35}$
D.~J.~James,$^{36}$
K.~Kuehn,$^{37,38}$
O.~Lahav,$^{4}$
S.~Lee,$^{39}$
J.~L.~Marshall,$^{40}$
J. Mena-Fern{\'a}ndez,$^{41}$
R.~Miquel,$^{42,26}$
J.~Muir,$^{43}$
J.~Myles,$^{44}$
A.~Palmese,$^{45}$
A.~A.~Plazas~Malag\'on,$^{46,47}$
A.~Porredon,$^{48,49}$
S.~Samuroff,$^{50,26}$
E.~Sanchez,$^{48}$
D.~Sanchez Cid,$^{48}$
I.~Sevilla-Noarbe,$^{48}$
E.~Suchyta,$^{51}$
G.~Tarle,$^{52}$
C.~To,$^{34}$
D.~L.~Tucker,$^{21}$
V.~Vikram,$^{22}$
A.~R.~Walker,$^{19}$
N.~Weaverdyck,$^{53,54}$
and J.~Weller$^{55,56}$\\
\begin{center} (DES Collaboration) \end{center}
}}
\vspace{0.1cm}

\date{Accepted XXX. Received YYY; in original form ZZZ}

\pubyear{2025}

\begin{document}

\label{firstpage}
\pagerange{\pageref{firstpage}--\pageref{lastpage}}
\AddToShipoutPictureBG*{%
  \AtPageUpperLeft{%
    \hspace{0.75\paperwidth}%
    \raisebox{-4.5\baselineskip}{%
      \makebox[0pt][l]{\textnormal{FERMILAB-PUB-24-0950-PPD}}
}}}%
\maketitle


\maketitle

\begin{abstract}
Recent cosmological analyses measuring distances of Type Ia Supernovae (SNe Ia) and Baryon Acoustic Oscillations (BAO) have all given similar hints at time-evolving dark energy. To examine whether underestimated SN~Ia systematics might be driving these results, \cite{George2024} compared overlapping SN events between \pplus\ and \DESVYR\ (20\% SNe are in common), and reported evidence for a $\sim$0.04 mag offset between the low and high-redshift distance measurements of this subsample of events. If these offsets are arbitrarily subtracted from the entire \DESVYR\ sample, the preference for evolving dark energy is reduced. In this paper, we reproduce this offset and show that it has two sources.
First, 43\% of the offset is due to \DESVYR\ improvements in the modelling of supernova intrinsic scatter and host galaxy properties. These are scientifically-motivated modelling updates implemented in \DESVYR\ and their associated uncertainties are captured within the \DESVYR\ systematic error budget. 
Even if the less accurate scatter model and host properties from \pplus\ are used instead, 
the \DESVYR\ evidence for evolving dark energy is only reduced from 3.9$\sigma$ to 3.3$\sigma$. 
Second, 38\% of the offset is due to a misleading comparison because different selection functions characterize the DES subsets included in \pplus\ and \DESVYR\, and therefore
individual SN distance measurements are \textit{expected} to be different because of 
different bias corrections. 
In conclusion, we confirm the validity of the published \DESVYR\ results.
\end{abstract}


%%%%%%%%%%%%%%%%% BODY OF PAPER %%%%%%%%%%%%%%%%%%

\section{Context}
The cosmological results from the Dark Energy Survey Supernova Program (DES-SN) have been published in two stages:
(1) a small {\it spectroscopically classified} subset of $\sim 200$ events
combined with ${\sim}100$ previously released low-$z$ events from the community
\citep[\DESIIIYR:][]{Brout2019_DES3YR,DES3YR}, and
(2) a more complete {\it photometrically classified} sample of 1600 events,
combined with ${\sim}190$ low-$z$ events from the community 
\citep[\DESVYR:][]{Vincenzi2024,DES5YR}.
In between these two efforts, a much larger {\it spectroscopically confirmed} sample
of 1,701 publicly released light-curves was used to publish cosmology constraints 
\citep[\pplus: ][]{Brout2022_pplus,Scolnic2022_pplus}.
While \pplus\ is technically not a DES-SN result, the \pplus\ and \DESVYR\ analyses
included a significant overlap of software \citep{SNANA,PIPPIN,COSMOSIS},
people and development, particularly for the 
intrinsic scatter modelling based on dust 
\citep{BS21,DUST2DUST}
and Beams with Bias Corrections \citep[BBC: ][]{KS17}.

\begin{figure*}
    \includegraphics[width=0.9\linewidth]{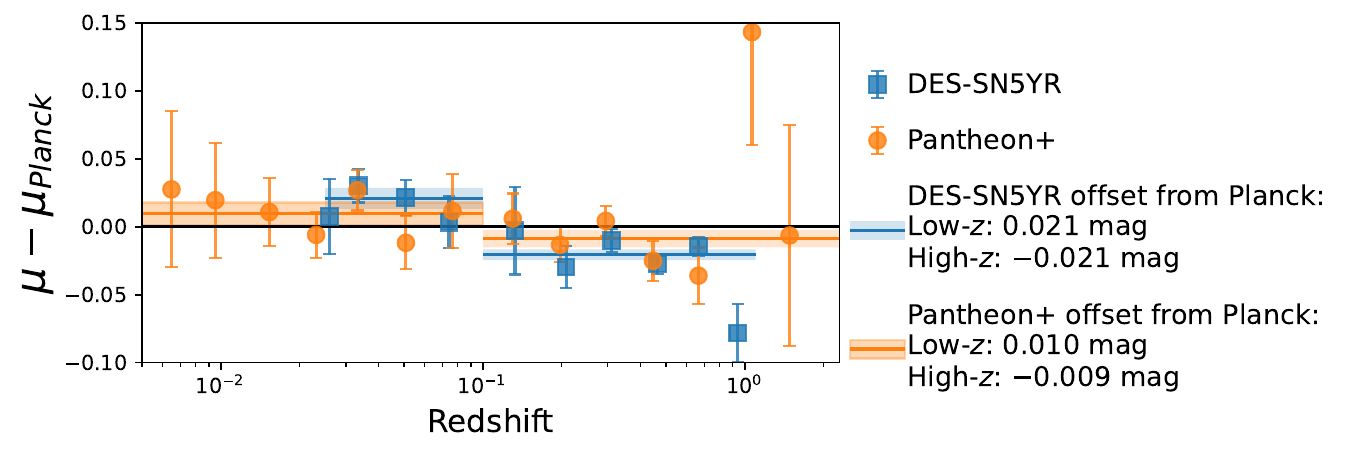}
    \caption{\pplus\ and \DESVYR\ binned Hubble residuals calculated w.r.t.\ a Flat$\Lambda$CDM cosmology assuming $\Omega_M = 0.315$ from \textit{Planck}.
    In each redshift bin we show the weighted mean of the Hubble residual and statistical-only uncertainties.
    The horizontal bands show the weighted mean of the Hubble residuals (and associated uncertainties) 
    above and below redshift 0.1 for both \pplus\ and \DESVYR.}
    \label{fig:hd_compare}
\end{figure*}

At low redshifts ($z<0.1$), there is significant sample overlap between \pplus\ and \DESVYR. 
More than 90\% of the low-$z$ SNe in \DESVYR\ are also included in \pplus\ 
\citep[184 low-$z$ overlapping events in total, 118 from the Foundation SN sample, 59 from CfA SN programs and 7 from the Carnegie Supernova Project;][]{2009ApJ...700..331H,2012ApJS..200...12H, 2017AJ....154..211K, Foley_Foundation}. \pplus\ includes $\sim500$ additional low-$z$ events that were not included in \DESVYR\ to minimize low-$z$ related systematics (in particular, calibration uncertainties). % end rk
At high redshifts ($z>0.1$), sample overlap between \pplus\ and \DESVYR\ includes 145 common DES SN events, which is $\sim$8\% of the \DESVYR\ high-$z$ sample and $\sim$14\% of the \pplus\ high-$z$ sample.

The binned \pplus\ and \DESVYR\ Hubble diagram residuals are compared in Fig.~\ref{fig:hd_compare} 
with respect to (w.r.t.) a
Flat$\Lambda$CDM cosmology model assuming the \textit{Planck} best-fit 
dark matter energy density $\Om = 0.315$ \citep{2020A&A...641A...6P}.
The binned Hubble residuals are consistent within 1$\sigma$ in every bin between \pplus\ and \DESVYR, and both datasets show a discrepancy w.r.t.\ \textit{Planck}. Discrepancies are in the same direction (positive residuals at low-$z$ and negative residuals at high-$z$) but for DES-SN5YR the effect is larger 
($\pm\sim$ 0.02 mag discrepancies for \DESVYR\ compared to $\pm\sim$0.01 mag discrepancies for \pplus) and uncertainties are smaller due to improved light curve modelling and larger high-$z$ statistics.
Using flat priors on cosmology parameters  
the ``SN-only" \pplus\ and \DESVYR\ analyses found consistent results in Flat$\Lambda$CDM ($\Omega_M= 0.334\pm0.018$ and $\Omega_M=0.352\pm 0.017$ respectively) and Flat$w$CDM \citep[$w=-0.90\pm0.14$ and $w=-0.80^{+0.14}_{-0.16}$ respectively, see also Fig.~12 in][]{DES5YR}. In Flat$w_0w_a$CDM,\footnote{Following the dark energy equation of state parametrization $w(a)=w_0+w_a(1-a)$.} the results are consistent to $\sim2\sigma$,\vspace{1mm}\\
\indent Pantheon+: $w_0=-0.93\pm0.15$, $w_a=-0.1^{+0.9}_{-2.0}$, \\
\indent DES SN5YR: $w_0=-0.36^{+0.36}_{-0.30}$, $w_a=-8.8^{+3.7}_{-4.5}$~.\vspace{1mm}\\
While the \pplus\ distance uncertainties are larger compared to \DESVYR, the \pplus\ cosmology uncertainties are smaller because of the (i) larger redshift range (0.001 to 2.26) compared to the \DESVYR\ redshift range (0.025 to 1.13), 
and (ii) additional systematic uncertainties included in \DESVYR.

Shortly after the \DESVYR\ constraints were published,
the Dark Energy Spectroscopic Instrument (DESI) Collaboration released cosmological results from the measurement of BAO in galaxy, quasar and Lyman-$\alpha$ forest tracers from the first year of observations \cite[DESI-BAO-Y1,][]{2024arXiv240403002D}. When combining DESI with CMB measurements \citep[CMB anisotropies from \textit{Planck} and CMB lensing data from \textit{Planck} and ACT, ][]{2022JCAP...09..039C,2020A&A...641A...1P}, a $2.6\sigma$ evidence for time-varying dark energy equation of state is found. This evidence is unchanged when the Pantheon+ SN constraints are added ($2.5\sigma$ evidence), but grows to 3.9$\sigma$ when \DESVYR\ constraints are used instead of Pantheon+.~These tantalizing hints for dynamical dark energy have motivated further scrutiny of these cosmological results, and re-analyses have focused on public data releases from \pplus\ and \DESVYR.

\newcommand{\UIII}{UNION3}
Other than DES-SN and Pantheon+, there was a another cosmological analysis using more than 2,000 spectroscopically confirmed SNe~Ia from public data releases:
the \UIII\ compilation \citep{UNION3_2023}.
For the Flat$\Lambda$CDM model, 
they found results consistent with \DESVYR\ and \pplus\ 
(best-fit $\Omega_M=0.356_{-0.026}^{+0.028}$). 
For the Flat$w$CDM model, 
they found $w=-0.74^{+0.17}_{-0.19}$,
a slightly larger deviation from a cosmological constant compared to \DESVYR\ and \pplus.
They do not publish SN-only results for the Flat$w_0w_a$CDM model.
When \UIII\ is combined with DESI BAO and CMB, there is a 3.1$\sigma$ evidence for time-evolving dark energy. 
The \UIII\ sample has a large overlap
of supernova light-curves with \pplus. 
However, the \UIII\ analysis used a Bayesian hierarchical framework 
\lq Unity\rq\ \citep{UNITY_2015}, 
which is very different from the methodology
used in \pplus\ and \DESVYR. \UIII\ has not released a Hubble diagram, and therefore we do not include 
\UIII\ comparisons in this investigation.

\section{Recent analysis by Efstathiou (2024)}

\citet{George2024} examined publicly released Hubble diagrams from \pplus\ and \DESVYR, and performed a
distance comparison with the overlapping SNe that are used in both samples. 
\citet{George2024} notes a $\sim$0.04 mag discrepancy in standardized brightnesses between the low-$z$ and high-$z$ overlapping events in \pplus\ and \DESVYR. He suggests that this discrepancy is the reason why the \DESVYR\ sample provides more significant evidence for evolving dark energy compared to \pplus\ and, in Flat$\Lambda$CDM, a larger $\Omega_M$ compared to CMB measurements from \textit{Planck}. 
The discrepancy highlighted by \citet{George2024} is measured using a very limited subsample of the data included in the two compilations 
($<20\%$ of the data, corresponding to only bright events at high-$z$).
Comparing the entire samples, however, the discrepancies are at the $~$0.01 mag level, at both high and low redshifts (Fig.~\ref{fig:hd_compare}).

In this paper, we do not address DESI results or discrepancies from \textit{Planck}, 
and in the spirit of blind analyses we do not comment on the [dis]agreement with respect to a cosmological constant. Our response addresses
the 0.04 mag discrepancy observed by \citet{George2024} between the \DESVYR\ and \pplus\, 
and we explain the reasons for this mismatch. 
We focus on the 118 low-$z$ Foundation events and 
145 high-$z$ DES events that overlap in \pplus\ and \DESVYR\ 
and consider the distance modulus offset:
\begin{equation}
\begin{split}
\dmuOffset=
\bigl\langle\mu_{\mathrm{\pplus}} - \mu_{\mathrm{\DESVYR}}\bigr\rangle_{\mathrm{Foundation}} &-\\
\bigl\langle\mu_{\mathrm{\pplus}} - \mu_{\mathrm{\DESVYR}}\bigr\rangle_{\mathrm{DES}}&
\end{split}
\label{eq:offset}
\end{equation}
where $\langle\mu_{\mathrm{\pplus}} - \mu_{\mathrm{\DESVYR}}\rangle$ is
the mean difference between \pplus\ and \DESVYR\ distance moduli ($\mu_{\mathrm{\pplus}}$ and $\mu_{\mathrm{\DESVYR}}$, respectively) computed from the inverse-variance weighted average over the overlapping Foundation or DES SNe.
Since we only use SNe Ia to measure \textit{relative} distances, any constant offset between the two datasets would be absorbed by a combination of the Hubble constant $H_0$ and SN Ia absolute magnitude $M_0$, and not the cosmological parameters of interest here. In other words, if the distance moduli from \pplus\ and \DESVYR\ differed by the same amount at both low and high redshifts, this would not affect the inferred cosmology.

We note that \citet{George2024} did not use SN distance moduli $\mu$, 
but instead defined SN magnitudes as $m = \mu - M_0$, with $M_0$ fixed to $-19.33$. 
Using this definition, \citet{George2024} found that,
\begin{eqnarray}
\bigl\langle m_{\mathrm{\pplus}} -  m_{\mathrm{\DESVYR}}\bigr\rangle_{\mathrm{Foundation}}\sim-0.05,~ \\
\bigl\langle m_{\mathrm{\pplus}} - m_{\mathrm{\DESVYR}}\bigr\rangle_{\mathrm{DES}}~~~~~~\sim-0.01,~
\end{eqnarray}
suggesting that the most significant discrepancies between \pplus\ and \DESVYR\ are in the analyses of the Foundation SN sample.
However, this approach implicitly assumes that $M_0$ is the same for \pplus\ and \DESVYR,
which is not true because $M_0$ depends on the light curve model training and analysis, which differ between 
these two analyses.
We therefore focus on the quantity $\dmuOffset$ (Eq.~\ref{eq:offset}) 
as it eliminates any confusion in the definition of $M_0$.

We provide a brief analysis recap in Sec.~\ref{sec:analysis_recap}.
The main reasons for the observed discrepancy is given in 
Sections~\ref{sec:analysis_updates} and \ref{sec:selection_effects},
and some concluding remarks are in Sec.~\ref{sec:conclude}.

\section{SNe Ia standardization and the role of bias corrections}
\label{sec:analysis_recap}
Here, we briefly summarize the main steps necessary to standardize SNe Ia brightnesses and correct for selection biases.
The first step to standardize SN brightnesses is to fit each SN light-curve for the parameters
$\{m_x,x_1,c\}$, which are the amplitude, stretch, and colour of a SN, respectively
\citep{Guy2007,Betoule2014}. For each SN, these parameters are used to measure its standardized apparent magnitude, $m_x$ \citep{Tripp1998}
\begin{equation}
\mxstd = m_x+\alpha x_1 -\beta c + \gamma G_{M_{\star}}
\label{eq:Tripp}
\end{equation}
where $\alpha$, $\beta$ and $\gamma$ are globally fitted nuisance parameters modelling the stretch-, colour- and host- luminosity dependencies respectively.
$G_{M_{\star}}$ is a $\pm1/2$ step function that describes the magnitude offset observed between SNe found in high stellar mass ($M_{\star}>10^{10}M_{\odot}$) and low stellar mass ($M_{\star}<10^{10}M_{\odot}$) galaxies (the so-called \lq mass step\rq).  
SN standardized apparent brightnesses are used to measure SN distance moduli $\mu$, defined as
\begin{equation}
\mu = \mxstd - \mubias - M_0
\label{eq:dist}
\end{equation}
where $M_0$ is the SN absolute brightness (fully degenerate with $H_0$). The term $\mubias$ corrects each SN distance for selection effects and analysis biases. $\mubias$ are determined from large simulations and are defined as $\mubias = \mu_{\rm sim} - \mu_{\rm true}$, where $\mu_{\rm sim} = \mxstd-M_0$ are the derived SN distances for simulated SN events and $\mu_{\rm true}$ are the true simulated SN distances. We discuss $\mubias$ in more detail below. 

\begin{figure}
    \includegraphics[width=\linewidth]{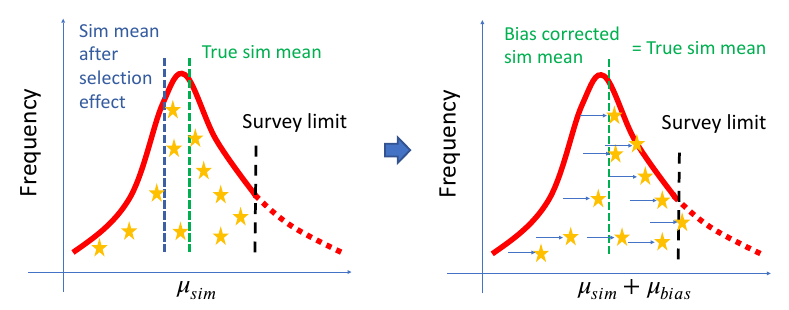}
    \caption{\textit{Left panel}: 
    Derived distances from simulated SN events $\mu_{\rm sim}$ for SNe in a given 
    4D bin of $z,x_1,c,M_{\star}$.
    The derived mean of the simulated data (blue vertical dashed line) is biased compared to the true simulated mean (green vertical dashed line). 
    \textit{Right panel}: 
   Simulation-derived bias corrections ($\mubias$) are added to \textit{every} SN Ia in the bin (horizontal arrows) such that the derived mean aligns with the true mean. The correction depends on the SN~Ia light curve model, scatter model, and instrumental noise.}
    \label{fig:biascorcartoon}
\end{figure}

\begin{table*}
\caption{Distance modulus offsets between Foundation (low-$z$) and DES (high-$z$) SNe common to \DESVYR\ and \pplus\ compilations.  The ``Contribution to $\Delta \mu_{\rm offset}$'' column shows the size of the change in $\Delta \mu_{\rm offset}$ due to each effect.  The ``Remaining $\mu_{\rm offset}$'' column shows the offset remaining after that effect has been reverted or removed.
}
	\centering
\begin{tabular}{lcc}
\specialrule{.12em}{.05em}{.05em} 
  & Contribution to & Remaining \\
 \textbf{Analysis changes applied to {\DESVYR}} & \textbf{$\dmuOffset$} [mag] & \textbf{$\dmuOffset$} [mag] \\
\specialrule{.12em}{.05em}{.05em} 
\hspace{0.05cm}None & &\textbf{$-$0.042} \\
\specialrule{.05em}{.05em}{.05em} 
\hspace{0.2cm} Revert to  Pantheon+ intrinsic scatter model $(*)$ & \textbf{0.008} &\textbf{$-$0.034} \\
\hspace{0.2cm} Revert to Pantheon+ host stellar mass estimations & \textbf{0.010} &\textbf{$-$0.024} \\
\specialrule{.05em}{.05em}{.05em} 
\hspace{0.2cm} Remove offset due to different selection functions ($\ddag$)& \textbf{0.016} & \textbf{$-$0.008}  \\
\specialrule{.12em}{.05em}{.05em} 
\end{tabular}
    \begin{itemize} \centering
    \item[]{$(*)$ Pantheon+ used the
    original BS21 intrinsic scatter model. 
    \DESVYR\ included this model in the systematic error budget, 
    but used an improved version of the BS21
    model for the nominal analysis.}
    \item[]{($\ddag$) This offset arises because the different DES subsamples have different corrections for selection functions.  See Section 5.}
   \end{itemize}
   \label{tab:mu_shifts}
\end{table*}

Most SN surveys are magnitude-limited. SNe Ia are intrinsically variable objects, with bluer ($c<0$) and longer duration ($x_1>0$) explosions being brighter than redder ($c<0$) and shorter duration ($x_1<0$) explosions. 
While the standardisation process (Eq.~\ref{eq:Tripp}) results in nearly uniform brightness,
the pre-standardized brightness variation is more than 1~mag. This variation would not lead to a bias in the  magnitudes $\mxstd$ in the hypothetical case that all SN Ia have exactly the same brightness after standardisation, without error. 
However, a ${\sim}15\%$ variation in brightness
remains after standardization, composed of remaining intrinsic variation in the 
explosions, measurement noise, and modelling error. 
Detected SNe~Ia have a selection bias that preferentially includes intrinsically brighter events near the magnitude limit of the survey (as defined by a selection function). 
This selection effect results in a distance bias that $\mubias$ in Eq.~\ref{eq:Tripp} corrects for; 
without this bias correction, the cosmology results would be significantly altered.

We illustrate this bias in Figure~\ref{fig:biascorcartoon}. 
The left panel represents the $\mu_{\rm sim}$ distribution of simulated SNe Ia in a given 4-dimensional bin of redshift, stretch, colour, and host stellar mass and only events brighter than the detection limit are observed.
The bias correction ($\mubias$)
adjusts all of the simulated SNe in the given bin such that the derived mean of $\mu_{\rm sim}$ matches the true mean, $\mu_{\rm true}$, that would be measured without selection bias.

In both the \DESVYR\ and \pplus\ analyses, BBC
models selection biases using large simulations and interpolates $\mubias$ in the 4-dimensional space mentioned above. 
In these large simulations, we model the survey cadence, instrument characteristics, survey selection effects and the true distribution of SN Ia magnitudes. 
The latter includes SN~Ia populations of stretch and colour to define a mean brightness model, 
as well as an ``intrinsic scatter'' model to account for brightness variations about the mean model.

The intrinsic distribution of SN Ia magnitudes is described by a physically-motivated model that captures features such as intrinsic colour variation and host galaxy dust.
This model is calibrated and fitted using data to reproduce
the observed distributions and correlations of SN Ia parameters. 
As more data have been collected, 
the scatter model has continuously improved. The first attempt at a model beyond a coherent mag shift at all epochs and wavelengths was a simple spectral-energy-distribution (SED) intrinsic scatter model
(\citealt{Kessler2013}, based on \citealt{G10} and \citealt{C11}), and it was used in the original Pantheon analysis \citep{Scolnic2018_Pantheon} and 
\DESIIIYR\ \citep{Brout2019_DES3YR}.
Next, a significant update based on the properties of dust in SN host galaxies \citep{BS21} was used in \pplus.
Finally, an updated version of the dust model \citep{DUST2DUST} 
was used in \DESVYR\ (see Section 4). 
The final bias correction applied to each SN Ia depends on the 
convolution of sample selection and the simulation of instrumental noise and intrinsic scatter.
\par

\section{Major analysis improvements between \pplus\ and \DESVYR}
\label{sec:analysis_updates}

The \DESVYR\ analysis introduced several improvements compared to the \pplus\ analysis, in terms of the light-curve fitting model (upgrading from SALT2 to SALT3), intrinsic scatter model used in bias corrections, estimates of host galaxy stellar masses, and photometric calibration (using the latest DES internal calibration). A comprehensive list of these improvements is presented and discussed in Appendix A. In this section, we discuss the two analysis updates that have a significant impact on $\dmuOffset$ and summarize our results in Table~\ref{tab:mu_shifts}. \vspace{0.1cm}\\

\noindent
\textbf{(1) SN Ia Intrinsic scatter model:} 
This model is essential to simulate and estimate accurate bias corrections on measured SN distances (Eq.~\ref{eq:dist} and Section 3). 
Both \pplus\ and \DESVYR\ use a model for scatter in which the colour of SN Ia is a combination of intrinsic variation and reddening by host-galaxy dust, which itself depends on host galaxy properties \citep{BS21}.
This model is described by
a set of 12 free parameters:
4 parameters characterizing intrinsic SN colour variations (mean and standard deviations of intrinsic colour and intrinsic $\beta$) + 8 parameters characterizing dust properties and their dependency on the host galaxy (mean and standard deviations of $R_V$ and dust extinction exponential coefficient $\tau$ in high and low mass galaxies). 

\pplus\ used the dust parameters originally presented in \citet{BS21}, 
which were manually tuned
such that the \SNANA\ simulation reproduces the trends observed in data. 
In \DESVYR, we improved the measurement of these 12 parameters using the forward-modelling fitting method in \citet{DUST2DUST}. 
The differences between the colour/dust parameters used in \pplus\ and \DESVYR\ 
are presented in Table~3 of \citet{Vincenzi2024}. 
Despite its overly simplistic approach to evaluating dust parameters, the original \pplus\ modelling approach was still considered plausible, 
and therefore included in the \DESVYR\ analysis 
as a potential source of systematic uncertainty.
In Figs.~13-15 and Table~8 in \citet{Vincenzi2024}, 
we present the effects of this source of systematic (labelled as \lq BS21\rq) and show that using the BS21 intrinsic scatter model (instead of the nominal model from \citealt{DUST2DUST}) introduces an offset of $\sim$0.010 mag between the low- and high-redshift samples of the \DESVYR\ Hubble diagram.
\vspace{0.005cm}\\

\noindent 
\textbf{(2) Host masses}: 
In \DESVYR, host-galaxy stellar masses have been updated for DES using deeper coadd photometry \citep{Wiseman2020_coadd}, and also updated for the low-$z$ samples (Foundation, CfA and CSP) to ensure that the same galaxy SED fitting method was used for both the 
high-$z$ and low-$z$ samples
(\S~2.5 in Vincenzi et al. 2024).
For the Foundation subset of \DESVYR, the host masses of 10 SNe Ia (out of 118 total common between \DESVYR and \pplus) were changed from low-mass galaxies to  high-mass galaxies ($>10^{10}M_{\odot}$), 
and vice-versa 
3 were changed from a high-mass
to low-mass galaxy ($<10^{10}M_{\odot}$).
For the DES subset, the host masses of 10 (out of 145 common between \DESVYR and \pplus) were changed to low-mass galaxies, and 2 were changed to high-mass galaxies. 
Comparing the common SNe, the average difference in host stellar mass between 
\pplus\ and \DESVYR\ is $-$0.16 dex for Foundation SN hosts and $+$0.07 dex for DES SN hosts. 
Neither analysis explicitly accounted for the host mass uncertainties, which results in a modest underestimate of systematics uncertainties.
\vspace{0.005cm}\\

Both analysis improvements contribute to the discrepancies highlighted by \citet{George2024}. 
To quantify the impact that the analysis improvements in \DESVYR\ would have on \pplus, 
one would ideally re-analyze \pplus\ sample using current codes and methods.
However, such a re-analysis on a 3-year old sample
is technically challenging on a short time-scale.
Instead, it is sufficient to take the more practical
approach of re-analyzing \DESVYR\ by substituting \pplus\ modeling choices.

In Table~\ref{tab:mu_shifts}, we show changes in $\dmuOffset$ when reverting the \DESVYR\ intrinsic scatter model and host mass values to match what was used in {\pplus}.
The observed $\dmuOffset$ is reduced from $-0.042$ mag to $-0.024$ mag 
(43\% reduction). 
The \DESVYR\ error budget adequately accounts for systematics related to the choice of intrinsic scatter model, but both \pplus\ and \DESVYR\ underestimated uncertainties due to host stellar masses. 
We have estimated the missing systematic uncertainty related to host stellar mass:
the total \DESVYR\ systematic error on $w$ (assuming Flat$w$CDM) 
and $w_0$ or $w_a$ (assuming Flat$w_0w_a$CDM) increase by less than 3\%.

Finally, we estimate how \DESVYR\ cosmological results would change when reverting to the 
older (less accurate)
scatter model and host properties from \pplus.
The \DESVYR\ evidence for evolving dark energy is mildly reduced from 3.9$\sigma$ to 3.3$\sigma$.

\subsection{A note on bias-corrected distances in \pplus\ and \DESVYR}
The updated light curve and intrinsic scatter models resulted in a constant shift of 0.04~mag 
between bias-corrections in \pplus\ and \DESVYR. 
While this 0.04~mag offset may confuse the interpretation of distances in the two analyses,
this offset \textit{has no effect on the cosmological results of \DESVYR\ or \pplus} because
SN dark energy constraints are insensitive to global offsets in the distances.

\section{Modelling Selection effects}
\label{sec:selection_effects}

In Section~4, we show the impact of explicit analysis changes between \pplus\ and \DESVYR. However, even if both analyses were identical in their methods and assumptions, we still expect to see a non-zero $\dmuOffset$ when performing a direct object-to-object comparison between the \pplus\ and \DESVYR\ data compilations. 
The reason for this expectation
is that the selection of the overlapping DES sub-sample embedded in \pplus\ and {\DESVYR} is very different. 
As noted in the previous two Sections, changes to the selection function also change the bias corrections.

\begin{figure}
    \includegraphics[width=\linewidth]{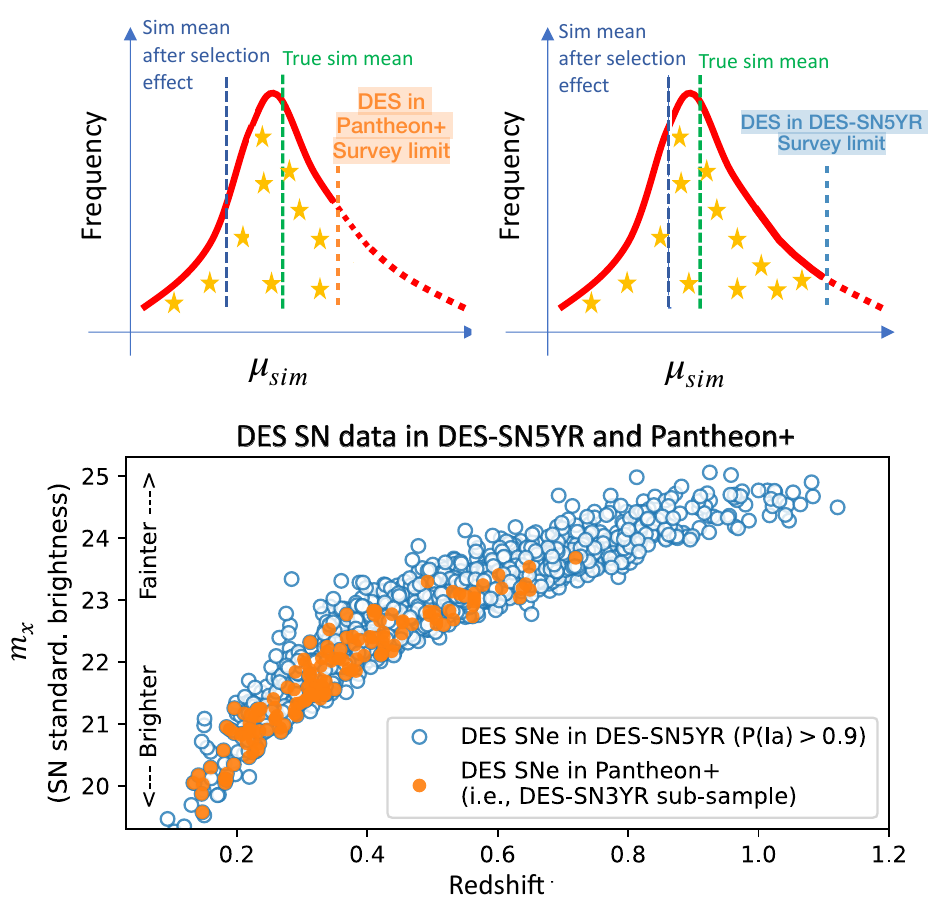}
    \caption{Illustration of the different selection functions characterizing the DES sub-samples in \DESVYR\ and \pplus. The \textit{upper-left} panel illustrates the DES SNe included in \pplus\, which are intentionally selected to be the brightest and highest signal-to-noise DES SNe, whereas the \textit{upper-right} panel shows \DESVYR\ with a significantly more complete sample. In the lower panel, we show SALT3-fitted $m_x$ \textit{vs} redshift for the real DES data in \DESVYR\ (empty blue circles) and in \pplus\ (filled orange circles). For \DESVYR, we only plot SNe with high ($>90\%$) probability of being type Ia. Even at lower redshifts ($z<0.4$), the sample of DES SNe in \pplus\ is not complete and clearly biased toward the brightest events.}
    \label{fig:brightness_redshfit}
\end{figure}

\begin{figure}
    \includegraphics[width=0.9\linewidth]{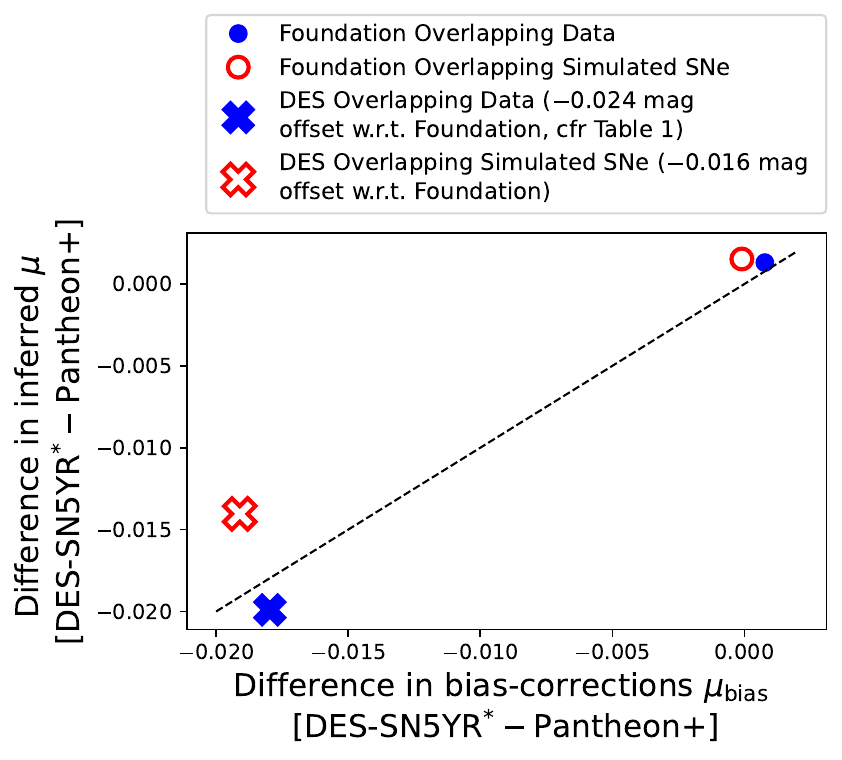}
    \caption{For the DES and Foundation sub-samples of overlapping SNe between \DESVYRstar\ and \pplus,
    we compare the difference in
    bias corrections $\mubias$ with the difference in distance moduli, $\mu$. 
    Differences relative to the Foundation (DES) sub-sample of overlapping data are marked with circles (crosses). Filled blue symbols correspond to differences in  the real data, while empty red symbols correspond to differences estimated from simulations. Instead of using the original \DESVYR\ distance moduli, we use distances from a re-analysis using the \pplus\ modelling as described in Section 4. For this reason, differences in measured distance moduli from the data are $-0.024$ (and not $-0.042$, see Table \ref{tab:mu_shifts}).
    To avoid confusion with different $M_0$, we subtract 0.04~mag from all the \DESVYR\ bias-corrections. 
    Using simulations, we quantify the expected differences in $\mu$ between the DES sub-samples in \pplus\ and \DESVYR\ to be $-0.016$.}
    \label{fig:mubias_color}
\end{figure}

The effects of bias corrections on \pplus\ and \DESVYR\ are illustrated in Fig.~\ref{fig:brightness_redshfit}. The \pplus\ analysis included the \DESIIIYR\ sub-sample, i.e., a sample of the \textit{brightest} DES SNe for which spectroscopic follow-up of the live transient was available (207 SNe). Therefore, this sample is characterized by \textit{strong selection biases} and requires \textit{large bias corrections}. In sharp contrast to this earlier analysis, the \DESVYR\ analysis used photometric classification \citep{2020MNRAS.491.4277M, SCONE_Qu, 2022MNRAS.514.5159M} and selected all DES SNe for which the host spectroscopic redshifts was available. 
This sub-sample is significantly deeper in redshift and more complete, and it includes a significantly larger number of SNe (1635 SNe). 

These fundamental differences in selection functions result in different bias corrections between the \pplus\ and \DESVYR\ analyses and make it difficult to compare the overlapping events as proposed in \citet{George2024}. 
Since we can accurately model the selection functions of the DES subsample included in \pplus\ and the \DESVYR\ sample, we can reproduce this effect using simulations. We generate a large ($\times10$) DES-like simulation and apply the \pplus\ DES selection function
and the \DESVYR\ selection function to obtain two samples that reflect the DES subsamples included in \pplus\ and \DESVYR.
We identify the overlapping simulated events between the two samples and compare their bias-corrections.

In Fig.~\ref{fig:mubias_color}, we compare
differences in bias corrections,
  $\mubias(\textrm{\DESVYRstar}) - \mubias(\textrm{\pplus})$
with differences in distance moduli $\mu(\textrm{\DESVYRstar}) - \mu(\textrm{\pplus})$.
For this comparison, \DESVYRstar\ is analyzed using \pplus\ modelling choices as described
in Section~4 and Table~\ref{tab:mu_shifts}. 
Using a consistent analysis avoids conflating multiple effects and more clearly shows the effect of bias corrections described in this section. As shown in Fig.~\ref{fig:mubias_color}, for the low-$z$ samples ($z<0.1$), the difference in $\mubias$ is negligible because
the sample selection is nearly the same for the two analyses. 
At higher redshifts ($z>0.1$), the data show large differences in $\mubias$ ($\sim0.02$) that are directly reflected into differences in $\mu$. 
These differences are
reproduced by our simulations at the 5~millimag level.
Our simulations show that differences in $\mubias$ are expected when analysing the two subsamples in the context of the (more biased towards brighter events) \pplus\ analysis or in the context of the (significantly more complete) \DESVYR\ analysis. 
In particular, both in data and simulations we find that \pplus\ bias corrections differ by approximately $\sim-$0.02 mag.

From our simulations, we estimate that this bias-correction effect contributes 
$\sim$0.016~mag to the $\dmuOffset$ observed by \citet{George2024}. However, we highlight that this effect is only important when performing an object-to-object comparison like the one presented by \citet{George2024}.
This effect is not relevant for the {\it full-sample}
Hubble diagram comparison shown in Fig.~\ref{fig:hd_compare} because each full sample is
corrected for their specific sample biases. In other words, in Fig.~\ref{fig:hd_compare} we compare the bias-corrected \lq Data mean\rq\ at each redshift bin (following the terminology of Fig.~\ref{fig:biascorcartoon}) rather than individual objects removed from their original context. 

\section{Conclusion}
\label{sec:conclude}
\citet{George2024} noted a 0.04~mag low-vs-high redshift distance offset (Eq.~\ref{eq:offset})
between overlapping \pplus\ and \DESVYR\ events. 
We have investigated this offset and find that it is explained as follow.
\begin{itemize}
    \item{\textbf{Two analysis improvements since \pplus}: These improvements are related to the intrinsic scatter model and host stellar mass estimates, and account for 0.018 mag discrepancy between \pplus\ and \DESVYR\ (from $-0.042$ to $-0.024$, see Table~\ref{tab:mu_shifts});} 
    \item{\textbf{Selection differences between \pplus\ and \DESVYR:} 
    Larger distance bias corrections are required for the more heavily biased \pplus\ sample of spectroscopically identified events, compared to smaller bias corrections for the more complete sample of photometrically classified events in \DESVYR\ (Fig.~\ref{fig:mubias_color}). This difference in selection functions does not affect cosmology results, but leads to misleading conclusions in an object-to-object comparison like the one presented by \citet{George2024}, where only 20\% of the brightest SNe are selected from both analyses. This effect account for an additional 0.016 mag discrepancy between \pplus\ and \DESVYR\ (from $-0.024$ to $-0.008$, see Table~\ref{tab:mu_shifts}).
    This biased comparison can be avoided by comparing the binned \pplus\ and \DESVYR\ Hubble diagrams 
    as shown in Fig.~\ref{fig:hd_compare}.}
\end{itemize}

Near the completion of our response to \citet{George2024},
\citet{Notari2024} performed a re-analysis of \pplus\ and \DESVYR\ in which
common SNe are excluded from one sample in order to perform a combined analysis 
of the two independent samples. We have not performed a detailed investigation of 
their analysis or claims, but we note a few qualitative issues with their analysis.
First, removing or adding events changes the selection criteria
and hence requires updated bias corrections. Second, while the 0.04~mag $M_0$ offset
between \pplus\ and \DESVYR\ distances does not impact cosmology results
from each separate analysis, this offset can bias results from combining
distances without a re-analysis using consistent SN~Ia modelling and bias corrections.

In conclusion, we hope that this post-publication investigation offers the community 
valuable insights into the \DESVYR\ and \pplus\ analyses, 
and it reinforces one of the main conclusions drawn from the \DESVYR\ analysis \citep{Vincenzi2024}: 
the limiting factor and largest source of systematics in current SN~Ia cosmology analysis is the modelling of intrinsic scatter and SN-host correlations in bias corrections. 
Nevertheless, these uncertainties are usually included in the error budget of the current data sets, and thus accounted for in the uncertainties on cosmological parameters. In particular, the \DESVYR\ analysis accounted for several additional sources of systematic uncertainties related to intrinsic scatter and SN-host correlations that have not been included in any previous analysis (including \pplus).
Here we identified one untracked systematic uncertainty due to host galaxy measurement uncertainties, which increases the previously reported uncertainties by about 3\%. 

Upcoming data from the Zwicky Transient Factory \citep{2024arXiv240904346R}, 
the Vera C. Rubin Observatory, and the Roman Space Telescope will improve our understanding and control of these sources of systematics.

\section*{Contribution Statement and Acknowledgments}
M.V. led the primary analysis and drafted the manuscript; R.K. and P.S. contributed substantially to the analysis, manuscript preparation, and figure curation. J.L. contributed to the analysis and provided comments on the manuscript. D.S. and T.M.D. advised on the analysis, provided detailed feedback on the manuscript, and served as internal reviewers. A.M., B.R., B.P., C.L., D.B., M. Sa., M.Sm.,  M.Su., P.W., L.G., R.C. and J.Mu. provided comments on the manuscript.
The remaining authors have made contributions to this paper that include, but are not limited to, the construction of DECam and other aspects of collecting the data; data processing and calibration; developing broadly used methods, codes, and simulations; running the pipelines and validation tests; and promoting the science analysis. This paper has gone through internal review by the DES collaboration.

Funding for the DES Projects has been provided by the U.S. Department of Energy, the U.S. National Science Foundation, the Ministry of Science and Education of Spain, 
the Science and Technology Facilities Council of the United Kingdom, the Higher Education Funding Council for England, the National Center for Supercomputing  Applications at the University of Illinois at Urbana-Champaign, the Kavli Institute of Cosmological Physics at the University of Chicago, 
the Center for Cosmology and Astro-Particle Physics at the Ohio State University, the Mitchell Institute for Fundamental Physics and Astronomy at Texas A\&M University, Financiadora de Estudos e Projetos, 
Funda{\c c}{\~a}o Carlos Chagas Filho de Amparo {\`a} Pesquisa do Estado do Rio de Janeiro, Conselho Nacional de Desenvolvimento Cient{\'i}fico e Tecnol{\'o}gico and 
the Minist{\'e}rio da Ci{\^e}ncia, Tecnologia e Inova{\c c}{\~a}o, the Deutsche Forschungsgemeinschaft and the Collaborating Institutions in the Dark Energy Survey. 

The Collaborating Institutions are Argonne National Laboratory, the University of California at Santa Cruz, the University of Cambridge, Centro de Investigaciones Energ{\'e}ticas, 
Medioambientales y Tecnol{\'o}gicas-Madrid, the University of Chicago, University College London, the DES-Brazil Consortium, the University of Edinburgh, 
the Eidgen{\"o}ssische Technische Hochschule (ETH) Z{\"u}rich, 
Fermi National Accelerator Laboratory, the University of Illinois at Urbana-Champaign, the Institut de Ci{\`e}ncies de l'Espai (IEEC/CSIC), 
the Institut de F{\'i}sica d'Altes Energies, Lawrence Berkeley National Laboratory, the Ludwig-Maximilians Universit{\"a}t M{\"u}nchen and the associated Excellence Cluster Universe, 
the University of Michigan, NSF's NOIRLab, the University of Nottingham, The Ohio State University, the University of Pennsylvania, the University of Portsmouth, 
SLAC National Accelerator Laboratory, Stanford University, the University of Sussex, Texas A\&M University, and the OzDES Membership Consortium.

R.K. is supported by DOE grant DE-SC0009924.  T.M.D.\ is the recipient of an Australian Research Council Laureate Fellowship (FL180100168) funded by the Australian Government. A.M.\ is supported by the ARC Discovery Early Career Researcher Award (DECRA) project number DE230100055. L.G. acknowledges financial support from AGAUR, CSIC, MCIN and AEI 10.13039/501100011033 under projects PID2023-151307NB-I00, PIE 20215AT016, CEX2020-001058-M, ILINK23001, COOPB2304, and 2021-SGR-01270. We acknowledge the University of Chicago’s Research Computing Center for their support of this work.

Based in part on observations at Cerro Tololo Inter-American Observatory at NSF's NOIRLab (NOIRLab Prop. ID 2012B-0001; PI: J. Frieman), which is managed by the Association of Universities for Research in Astronomy (AURA) under a cooperative agreement with the National Science Foundation. Based in part on data acquired at the Anglo-Australian Telescope. We acknowledge the traditional custodians of the land on which the AAT stands, the Gamilaraay people, and pay our respects to elders past and present.

The DES data management system is supported by the National Science Foundation under Grant Numbers AST-1138766 and AST-1536171.
The DES participants from Spanish institutions are partially supported by MICINN under grants ESP2017-89838, PGC2018-094773, PGC2018-102021, SEV-2016-0588, SEV-2016-0597, and MDM-2015-0509, some of which include ERDF funds from the European Union. IFAE is partially funded by the CERCA program of the Generalitat de Catalunya.
Research leading to these results has received funding from the European Research
Council under the European Union's Seventh Framework Program (FP7/2007-2013) including ERC grant agreements 240672, 291329, and 306478.
We  acknowledge support from the Brazilian Instituto Nacional de Ci\^encia
e Tecnologia (INCT) do e-Universo (CNPq grant 465376/2014-2).
This manuscript has been authored by Fermi Research Alliance, LLC under Contract No. DE-AC02-07CH11359 with the U.S. Department of Energy, Office of Science, Office of High Energy Physics.

%%%%%%%%%%%%%%%%%%%%%%%%%%%%%%%%%%%%%%%%%%%%%%%%%%
\section*{Data Availability}

All data used in this analysis are publicly available at \url{https://github.com/PantheonPlusSH0ES/DataRelease} and \url{https://github.com/des-science/DES-SN5YR}.


\bibliographystyle{mnras}
\bibliography{main}

%%%%%%%%%%%%%%%%% APPENDICES %%%%%%%%%%%%%%%%%%%%%

\appendix
\section*{Appendix: Additional analysis improvements in \DESVYR\ from \pplus\ }

In this Appendix, we list the additional analysis changes introduced in \DESVYR\ compared to \pplus. In contrast to the analysis updates discussed in Section~4, 
the updates discussed here have a negligible effect on the $\dmuOffset$ discussed by \citet{George2024}.\vspace{0.01cm}\\

\textbf{Light-curve fitting model:} 
\pplus\ used the SALT2 light-curve model from \citet{Betoule2014}
to fit each event for the standardizaton parameters
$\{m_x,x_1,c\}$.
In \DESVYR, we upgraded to the recently published SALT3 model \citep{K21_SALT3}, after carefully testing the SALT2 and SALT3 models on published cosmological samples \citep{Taylor2023}. 

\begin{figure}
    \includegraphics[width=0.95\linewidth]{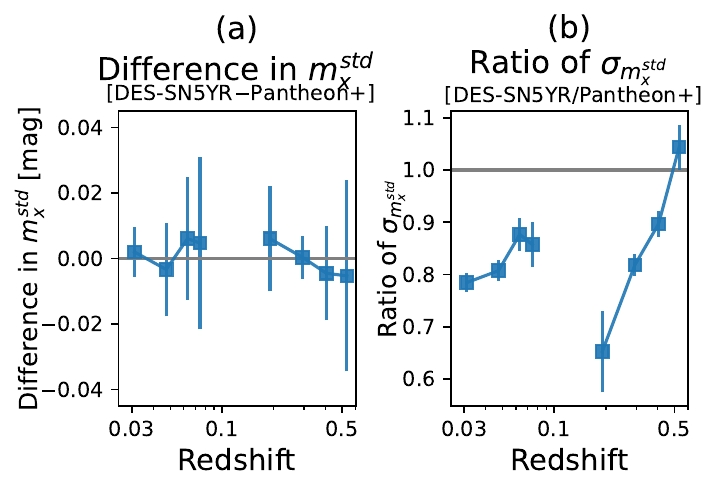}
    \caption{\textit{Left panel}: 
    Median $\mxstd$ difference vs. redshift (and error on the median) 
    for common SNe in \DESVYR\ and \pplus. 
    Across all redshifts, differences in $\mxstd$ are consistent with zero. 
    \textit{Right panel}: ratio of $\mxstd$ uncertainties between \DESVYR\ and \pplus. 
    For a more direct comparison,  we measure 
    $\mxstd$ and its uncertainty
    using the same nuisance parameters for both \DESVYR\ and \pplus: $\alpha=0.15$, $\beta=3.1$, $\gamma=0$.}    
    \label{fig:tripp}
\end{figure}

We compare the SALT2 and SALT3 standardized
$\mxstd$ in Fig.~\ref{fig:tripp}a and find
that differences are negligible.
Therefore, we do not expect the differences between SALT2 and SALT3 models to explain 
any of the trends highlighted by \citet{George2024}.

While the $\mxstd$ differences are small, its uncertainty is
significantly reduced in \DESVYR\ (see Fig.~\ref{fig:tripp}b). 
This reduction is due to the broader wavelength range covered by the SALT3 model. The SALT3 mean rest-frame wavelength per band extends 1000~\AA\ farther into the near-infrared compared to SALT2 (from 7000~\AA\ to 8000~\AA), enabling the use of Foundation and DES $z$-band data at low redshifts (where observed $z$-band correspond to $\sim$8000~\AA\ in the rest-frame). The SALT3 model therefore results in larger weight given to low-$z$ SNe in \DESVYR\ compared to \pplus. \vspace{0.01cm}

\noindent
\textbf{Calibration:}
Photometric calibration for both \pplus\ and \DESVYR\ is presented by \citep{BroutFragilistic} 
using a cross-calibration approach. 
For the Foundation SN sample, the same calibration was used for both \pplus\ and \DESVYR.
However, the DES subsets in \pplus and \DESVYR\ are treated independently because each sample was reprocessed independently. 
The same Scene Modelling Photometry code \citep[][]{SMP_2019, 2024arXiv240605046S} was used to produce light curves for both sub-samples, but the DES survey calibration was updated towards the end of the survey \citep[compare][]{2018AJ....155...41B, Sevilla_Noarbe_2021}
and therefore \DESVYR\ required slightly different AB calibration offsets 
($-$0.004 to 0.009 mag) compared to the earlier DES-SN3YR data release included in \pplus. 
The final AB calibration offsets used in \pplus\ and \DESVYR\ are presented in Table~3 of \citet{BroutFragilistic} and their uncertainties propagated in our systematic error budget. 
These calibration offsets are significantly smaller than the observed $\dmuOffset$ (Eq.~\ref{eq:offset}) 
and are therefore unlikely to have much impact on it. \vspace{0.01cm}\\

\noindent
\textbf{Nuisance parameters:} 
\pplus\ and \DESVYR\ each used their BBC-fitted
nuisance parameters for SN standardization and distance estimates: \pplus\ found $\alpha$=0.148, $\beta$=3.09 and $\gamma$=0, while \DESVYR\ found $\alpha$=0.161, $\beta$=3.12 and $\gamma$=0.038. Therefore, when comparing distances for the same events across the two analyses, differences are expected.
\vspace{0.01cm}\\

\noindent
\textbf{Beams with Bias Corrections:} 
the BBC framework was used both in \pplus\ and \DESVYR, and there have been several code updates\footnote{\URLBBC} between the publication of \pplus\ and \DESVYR. 
One of the most significant code updates
was related to the treatment of the parameter $\beta$ in the calculation of the bias-corrections. 
In \pplus, the (wrong) \textit{intrinsic} $\beta$ parameter 
($\beta_{\mathrm{intr}}{\sim}2$) 
was used in Eq.~\ref{eq:Tripp}
to estimate bias corrections, instead of the effective 
$\beta$ ($\beta_{\mathrm{eff}}{\sim}3$), which is a combination of intrinsic $\beta$ and extrinsic dust law.

We do not test this by reverting the entire \DESVYR\ to use the same SALT2 model as \pplus\ because Fig.~\ref{fig:tripp} shows that discrepancies are negligible. We also do not recompute the \DESVYR\ calibration because differences between calibration offsets applied to the DES sub-samples in \pplus\ and \DESVYR\ are also significantly smaller than $\dmuOffset$. 

Reprocessing \DESVYR\ using the (obsolete) \pplus\ version of the 
BBC code does not have a significant effect on $\dmuOffset$. 
However, reprocessing the full Pantheon+ sample (not just the overlapping sample) using the current BBC code results in an average distance change of
${\sim} {+0.005}$~mag at low-$z$ ($z<0.1$) and 
${\sim} {-0.005}$~mag at high-$z$ ($z > 0.1$). 
Therefore, BBC code updates impact the overall Pantheon+ Hubble diagram. 
As a final test, we reprocessed
the \DESVYR\ analysis (which was frozen nearly 1.5 years ago) with the current BBC code and found a distance change of
${\sim}0.001$~mag between low-$z$ and high-$z$.
\section*{Affiliations}
\onecolumn
\parbox{\textwidth}{
\footnotesize
$^{1}$ Department of Physics, University of Oxford, Denys Wilkinson Building, Keble Road, Oxford OX1 3RH, United Kingdom\\
$^{2}$ Department of Astronomy and Astrophysics, University of Chicago, Chicago, IL 60637, USA\\
$^{3}$ Kavli Institute for Cosmological Physics, University of Chicago, Chicago, IL 60637, USA\\
$^{4}$ Department of Physics \& Astronomy, University College London, Gower Street, London, WC1E 6BT, UK\\
$^{5}$ Department of Physics and Astronomy, University of Pennsylvania, Philadelphia, PA 19104, USA\\
$^{6}$ School of Mathematics and Physics, University of Queensland,  Brisbane, QLD 4072, Australia\\
$^{7}$ Department of Physics, Duke University Durham, NC 27708, USA\\
$^{8}$ Boston University Department of Astronomy, 725 Commonwealth Ave, Boston USA\\
$^{9}$ Institute of Space Sciences (ICE, CSIC),  Campus UAB, Carrer de Can Magrans, s/n,  08193 Barcelona, Spain\\
$^{10}$ Institut d'Estudis Espacials de Catalunya (IEEC), 08034 Barcelona, Spain\\
$^{11}$ Centre for Gravitational Astrophysics, College of Science, The Australian National University, ACT 2601, Australia\\
$^{12}$ The Research School of Astronomy and Astrophysics, Australian National University, ACT 2601, Australia\\
$^{13}$ Centre for Astrophysics \& Supercomputing, Swinburne University of Technology, Victoria 3122, Australia\\
$^{14}$ Univ Lyon, Univ Claude Bernard Lyon 1, CNRS, IP2I Lyon / IN2P3, IMR 5822, F-69622, Villeurbanne, France\\
$^{15}$ Department of Physics and Astronomy, Baylor University, Waco, TX 76706, USA\\
$^{16}$ Aix Marseille Univ, CNRS/IN2P3, CPPM, Marseille, France\\
$^{17}$ Physics Department, Lancaster University, Lancaster, LA1 4YB, UK\\
$^{18}$ School of Physics and Astronomy, University of Southampton,  Southampton, SO17 1BJ, UK\\
$^{19}$ Cerro Tololo Inter-American Observatory, NSF's National Optical-Infrared Astronomy Research Laboratory, Casilla 603, La Serena, Chile\\
$^{20}$ Laborat\'orio Interinstitucional de e-Astronomia - LIneA, Rua Gal. Jos\'e Cristino 77, Rio de Janeiro, RJ - 20921-400, Brazil\\
$^{21}$ Fermi National Accelerator Laboratory, P. O. Box 500, Batavia, IL 60510, USA\\
$^{22}$ Physik-Institut, University of Zurich, Winterthurerstrasse 190, CH-8057
Zurich, Switzerland\\
$^{23}$ University Observatory, Faculty of Physics, Ludwig-Maximilians-Universit\"at, Scheinerstr. 1, 81679 Munich, Germany\\
$^{24}$ Instituto de Astrofisica de Canarias, E-38205 La Laguna, Tenerife, Spain\\
$^{25}$ Universidad de La Laguna, Dpto. Astrofísica, E-38206 La Laguna, Tenerife, Spain\\
$^{26}$ Institut de F\'{\i}sica d'Altes Energies (IFAE), The Barcelona Institute of Science and Technology, Campus UAB, 08193 Bellaterra (Barcelona) Spain\\
$^{27}$ Hamburger Sternwarte, Universit\"{a}t Hamburg, Gojenbergsweg 112, 21029 Hamburg, Germany\\
$^{28}$ California Institute of Technology, 1200 East California Blvd, MC 249-17, Pasadena, CA 91125, USA\\
$^{29}$ Instituto de Fisica Teorica UAM/CSIC, Universidad Autonoma de Madrid, 28049 Madrid, Spain\\
$^{30}$ Institute of Cosmology and Gravitation, University of Portsmouth, Portsmouth, PO1 3FX, UK\\
$^{31}$ Center for Astrophysical Surveys, National Center for Supercomputing Applications, 1205 West Clark St., Urbana, IL 61801, USA\\
$^{32}$ Department of Astronomy, University of Illinois at Urbana-Champaign, 1002 W. Green Street, Urbana, IL 61801, USA\\
$^{33}$ Santa Cruz Institute for Particle Physics, Santa Cruz, CA 95064, USA\\
$^{34}$ Center for Cosmology and Astro-Particle Physics, The Ohio State University, Columbus, OH 43210, USA\\
$^{35}$ Department of Physics, The Ohio State University, Columbus, OH 43210, USA\\
$^{36}$ Center for Astrophysics $\vert$ Harvard \& Smithsonian, 60 Garden Street, Cambridge, MA 02138, USA\\
$^{37}$ Australian Astronomical Optics, Macquarie University, North Ryde, NSW 2113, Australia\\
$^{38}$ Lowell Observatory, 1400 Mars Hill Rd, Flagstaff, AZ 86001, USA\\
$^{39}$ Jet Propulsion Laboratory, California Institute of Technology, 4800 Oak Grove Dr., Pasadena, CA 91109, USA\\
$^{40}$ George P. and Cynthia Woods Mitchell Institute for Fundamental Physics and Astronomy, and Department of Physics and Astronomy, Texas A\&M University, College Station, TX 77843,  USA\\
$^{41}$ LPSC Grenoble - 53, Avenue des Martyrs 38026 Grenoble, France\\
$^{42}$ Instituci\'o Catalana de Recerca i Estudis Avan\c{c}ats, E-08010 Barcelona, Spain\\
$^{43}$ Perimeter Institute for Theoretical Physics, 31 Caroline St. North, Waterloo, ON N2L 2Y5, Canada\\
$^{44}$ Department of Astrophysical Sciences, Princeton University, Peyton Hall, Princeton, NJ 08544, USA\\
$^{45}$ Department of Physics, Carnegie Mellon University, Pittsburgh, Pennsylvania 15312, USA\\
$^{46}$ Kavli Institute for Particle Astrophysics \& Cosmology, P. O. Box 2450, Stanford University, Stanford, CA 94305, USA\\
$^{47}$ SLAC National Accelerator Laboratory, Menlo Park, CA 94025, USA\\
$^{48}$ Centro de Investigaciones Energ\'eticas, Medioambientales y Tecnol\'ogicas (CIEMAT), Madrid, Spain\\
$^{49}$ Ruhr University Bochum, Faculty of Physics and Astronomy, Astronomical Institute, German Centre for Cosmological Lensing, 44780 Bochum, Germany\\
$^{50}$ Department of Physics, Northeastern University, Boston, MA 02115, USA\\
$^{51}$ Computer Science and Mathematics Division, Oak Ridge National Laboratory, Oak Ridge, TN 37831\\
$^{52}$ Department of Physics, University of Michigan, Ann Arbor, MI 48109, USA\\
$^{53}$ Department of Astronomy, University of California, Berkeley,  501 Campbell Hall, Berkeley, CA 94720, USA\\
$^{54}$ Lawrence Berkeley National Laboratory, 1 Cyclotron Road, Berkeley, CA 94720, USA\\
$^{55}$ Max Planck Institute for Extraterrestrial Physics, Giessenbachstrasse, 85748 Garching, Germany\\
$^{56}$ Universit\"ats-Sternwarte, Fakult\"at f\"ur Physik, Ludwig-Maximilians Universit\"at M\"unchen, Scheinerstr. 1, 81679 M\"unchen, Germany\\}

%%%%%%%%%%%%%%%%%%%%%%%%%%%%%%%%%%%%%%%%%%%%%%%%%%

% Don't change these lines
\bsp	% typesetting comment
\label{lastpage}
\end{document}